# MultiStructRNA: a Python package for multi-algorithm RNA secondary structure prediction, ensemble analysis, and visualization

Yashrajsinh Jadeja[1], Haining Lin[1], Mihir Metkar[1*]

[1]Moderna, Inc., Cambridge, MA 02142, USA.

**[*]To whom correspondence should be addressed.**
E-mail: Mihir.Metkar@modernatx.com

## Abstract

***Summary:*** We introduce **MultiStructRNA**, a unified Python toolkit for RNA secondary structure prediction, ensemble analysis, and visualization. Although RNA secondary structure is central to RNA biology and therapeutic design, practical adoption is often hindered by fragmented tooling, incompatible input and output formats, and limited visualization support. *MultiStructRNA* addresses these challenges through a single high-level API that orchestrates multiple prediction algorithms, harmonizes results into a consistent schema, and provides reproducible, ensemble-aware metrics through an object model suited to both interactive notebooks and production pipelines. *MultiStructRNA* enables seamless switching between prediction methods without requiring workflow changes and supports both in-notebook and exportable visualizations. Designed for scalability, it supports high-throughput analyses and simplifies comparison across methods while standardizing downstream feature extraction. The current release also includes optional agent-readable workflow recipes that document dependency setup, backend-adapter conventions, SHAPE-data reconciliation, structure interpretation, and comparative sequence analyses. By integrating diverse RNA secondary structure packages within a common framework, *MultiStructRNA* streamlines structure analysis and facilitates its use in RNA design, optimization, and machine learning workflows.

***Availability and implementation:*** Availability and implementation: MIT-licensed source code, documentation, example data, tutorial notebooks, and optional agent-readable workflow recipes are available at https://github.com/modernatx/MultiStructRNA (Python >=3.10, optional external predictors installed separately).

## 1 Introduction

RNA secondary structure is a key determinant of RNA function. It governs RNA stability, localization, and interactions with proteins (Bose, Saleem, and Mustoe 2024, Cao *et al.* 2024). In basic biology, structure influences translation efficiency (Mauger *et al.* 2019) and intracellular trafficking (Das *et al.* 2021); in therapeutics, it contributes to RNA stability (Leppek *et al.* 2022), specificity and regulatory activity (Pardi *et al.* 2018). Thus, RNA structure plays a central role across both mechanistic studies and therapeutic design (Metkar, Pepin, and Moore 2024).

Over decades of research, a range of computational tools, based on both physics-based models and machine learning have been developed to predict RNA secondary structure (Sato and Hamada 2023). These tools support both fundamental analysis and rational RNA engineering. Structure prediction is now routinely used in research and applied settings, including the design of stable mRNAs (Leppek *et al.* 2022), guide RNAs (Kocak *et al.* 2019), and elements such as internal ribosome entry sites (IRES) that modulate translation initiation (Marques, Lacerda, and Romão 2022).

Many RNA structure-based analyses rely on multiple publicly available tools, each with its own strengths and limitations. While this diversity allows researchers to tailor predictions to specific applications, it creates practical challenges. Integrating tools into workflows often requires ad hoc scripting, reconciling incompatible input/output formats, and working around limited visualization capabilities. Additionally, some structural metrics are absent or inconsistently reported across tools, making comparisons or tool-switching difficult.

Meanwhile, Python's widespread adoption in bioinformatics, data science, and machine learning has made it a preferred environment for developing and executing these analyses. Practical workflows frequently compare predictions across algorithms and extract structural features for downstream applications. In particular, secondary structure features serve as key objective functions in sequence design and model optimization, where comparable metrics are essential for evaluating or substituting prediction methods (Zhang *et al.* 2023).

*MultiStructRNA* addresses these challenges through a unified Python package that integrates multiple RNA secondary structure prediction tools under a single high-level, object-oriented API. It eliminates the need for ad hoc scripting and tool-specific code by harmonizing outputs into a consistent data model, computing ensemble-aware metrics, and generating a range of visualizations. The package is also designed to be agent-native, with optional agent-readable workflow support that enables reproducible setup, backend integration, structure interpretation, and comparative analyses in human- and agent-assisted development environments. This design establishes a consistent and scalable foundation for RNA structure analysis, empowering researchers to move seamlessly between methods, extract meaningful features, and integrate structural insights into experimental and computational workflows.

With just a few function calls, users can perform structure prediction and feature extraction in Jupyter notebooks or incorporate them into larger bioinformatics pipelines. Harmonized metrics can be readily used as objective functions in multi-objective optimization or loss functions in machine learning models. Built as a subclass of Biopython's widely used SeqRecord object (Cock *et al.* 2009), *MultiStructRNA* also inherits a rich set of utilities for sequence manipulation and metadata management, making it both powerful and easy to adopt. In doing so, *MultiStructRNA* lays the groundwork for more accessible, reproducible, and integrative RNA structure analysis across research and therapeutic design. Its repository-level workflow documentation further supports reproducible use and contributor onboarding in increasingly agent-assisted scientific programming environments.

## 2 Approach

**Unified backends.** *MultiStructRNA* wraps several widely used RNA secondary structure prediction engines behind a common interface, including LinearFold (beam-search MFE) (Huang *et al.* 2019), LinearPartition (approximate partition function) (Zhang *et al.* 2020), ViennaRNA (MFE and partition function) (Lorenz *et al.* 2011), RNAstructure (MFE and partition function) (Reuter and Mathews 2010), and MXfold2 (machine-learning-based prediction) (Sato, Akiyama, and Sakakibara 2021). It also integrates visualization and analysis tools, including VARNA (Darty, Denise, and Ponty 2009), RNAvigate (Irving and Weeks 2024), and RiboGraphViz (DasLab 2025; https://github.com/DasLab/RiboGraphViz), enabling secondary structure plots, arc plots, and graph-based maximum ladder distance (MLD) calculations. Users can specify the RNA sequence, algorithm, and output directory, and optionally retain or discard intermediate data structures as object attributes to trade off speed and memory footprint while still exposing tool-specific parameters and visualization options.

**High-level API.** *MultiStructRNA* exposes a high-level interface for structure prediction and visualization from a single sequence object. Users call predict() to run one or more supported engines and plot() to render secondary-structure or arc plots derived from base-pairing probabilities. The predict() method also accepts a curated set of algorithm-specific parameters for additional control while preserving a consistent workflow. Under the hood, *MultiStructRNA* follows a three-layer architecture: a user-facing API layer, an orchestration layer that standardizes inputs/outputs and manages caching, and a backend “engine” layer that wraps external predictors.

**Standardization of input and output data.** MultiStructRNA normalizes outputs - e.g., dot-bracket structures, energy terms, base-pairing probability matrices, and ensemble-derived features - into a shared schema, enabling consistent downstream analysis regardless of the underlying prediction algorithm.

**Agent-readable workflow layer.** MultiStructRNA also includes optional repository-scoped workflow recipes that make common package tasks easier to execute reproducibly in both human and agent-assisted development environments. These recipes are not runtime dependencies. Instead, they encode package-specific procedures for dependency setup, backend-adapter conventions, interpretation of ensemble metrics, reconciliation of SHAPE-guided and unguided predictions, and comparative analysis of sequence variants. This layer translates the package architecture into repeatable workflows while preserving the common output schema.

**Ensemble-aware metrics and standard outputs.** In addition to standard outputs such as minimum free energy (MFE), ensemble free energy, dot-bracket representations of MFE or MEA (maximum expected accuracy) structures, and base-pairing probability matrices, MultiStructRNA computes ensemble-derived features from base-pairing probabilities. These include per-nucleotide unpaired probability, Shannon entropy, maximum ladder distance (MLD), and base-pairing log-odds together with its sequence-level average (average log-odds, ALO), returned both as global summary metrics and position-resolved vectors. The base-pairing log-odds transforms each per-position base-pairing probability $p$ into $\ln(p/(1 - p))$, expanding the dynamic range at the extremes of the 0-to-1 probability scale to provide fine-scale, orthogonal structural information that complements global metrics for downstream applications such as mRNA in-solution stability prediction(Yi *et al.* 2026). Results are stored as object attributes to support batch workflows and to enable explicit control over memory usage in large-scale analyses.

**Visualization.** MultiStructRNA generates secondary-structure plots via the plot() method, with options to highlight specific nucleotides or ranges and to color bases by their summed base-pairing probability. Arc plots that visualize base-pairing probabilities and interactions are enabled through RNAvigate (Irving and Weeks 2024) with arcs optionally colored by per-nucleotide pairing probability. Plots can be displayed in-notebook and/or exported as PNG images.

# 3 Implementation

**Python implementation.** MultiStructRNA is implemented in Python and is fully type annotated. External structure prediction tools are installed separately; MultiStructRNA detects available backends at runtime and provides clear errors or fallbacks when optional dependencies are unavailable. The package is tested on Python ≥3.10 on Ubuntu 22.04 (x86_64); compatibility with other operating systems and CPU architectures depends on the availability of the underlying third-party tools.

**Developer and agent-facing documentation.** The repository includes conventional contributor documentation together with five task-specific workflow recipes: multistructrna-setup, engine-integration, structure-interpretation, SHAPE-reconciliation, and comparative-analysis. Each recipe provides trigger contexts, expected inputs, stepwise procedures, and validation checks. For example, engine-integration records the backend-adapter contract and expected return schema, while shape-reconciliation documents guided versus unguided prediction and SHAPE-reactivity agreement workflows.

## API example

```
from multistructrna import MSRNA

rna_sequence = "AUGGCGUAGUUUACCGGUAUGCGAU"
rna_structure = MSRNA(seq=rna_sequence, name="test")
rna_structure.predict(
algorithm="linearpartition",
keep_summed_probability_vector=True,
)
rna_structure.plot(
plot_type="secondary_structure",
output_dir="./",
show=True,
hl_bpp=rna_structure.summed_probability_vector,
)
```

Additional tutorial notebooks, example commands and reproducible environment files are provided in the GitHub repository (https://github.com/modernatx/MultiStructRNA).

## 4 Results and use cases

Figure 1 summarizes the MultiStructRNA design philosophy and representative outputs.

- **Small-scale computation.** For one-off analyses, MultiStructRNA supports comparing MFE and MEA structures across algorithms, visualizing motifs of interest using secondary-structure and arc plots, and incorporating experimental data (e.g., SHAPE) to guide structure prediction where supported by underlying backends.
- **Large-scale computation.** For high-throughput secondary-structure analyses on HPC environments, MultiStructRNA supports parallel execution and efficient feature extraction. Predicted structure-derived metrics and base-pairing probabilities can be used as objective functions for RNA sequence and oligonucleotide design and as inputs for machine learning workflows.
- **Developer and agent-assisted workflows.** For contributors and advanced users, the workflow recipes support repeatable setup, troubleshooting, interpretation, and analysis workflows, including comparing predictions, checking whether SHAPE-guided predictions alter inferred structure, and producing consistent summaries of ensemble-derived metrics.

## 5 Limitations and future work

MultiStructRNA relies on external predictors; future work will explore packaging redistributable binaries for these predictors, pending tool authors' permissions. The current release focuses on a standard set of inputs and output options most commonly used or directly relevant for machine learning and optimization workflows. MultiStructRNA currently supports five popular external prediction engines: LinearFold, LinearPartition, ViennaRNA, RNAstructure, and MXfold2. Future work will expand the set of tool-specific parameters to fully leverage capabilities exposed by underlying tools, add support for additional predictors and modified nucleotide chemistries, and broaden benchmarking and documentation. As an open-source toolkit, MultiStructRNA welcomes contributions from the broader scientific community.

## 6 Conclusion

MultiStructRNA is a Python-based, publicly available toolkit that unifies multi-algorithmic structure prediction, ensemble-aware metrics, and visualization into a single, reproducible framework. By standardizing outputs and streamlining integration, it supports robust RNA structure analysis and provides an extensible foundation for community-driven development.

## Funding

This work was supported by Moderna, Inc.

## Supplementary data

Supplementary data are available at Bioinformatics online.

## Acknowledgements

We thank Kendall Loh for providing feedback and reviewing the manuscript, and Eric Ma, Sara Ali, and Mary Richardson for feedback on API design, documentation, and package usability.

## Conflict of interest

Y.J., H.L. and M.M. are current employees and shareholders of Moderna, Inc.

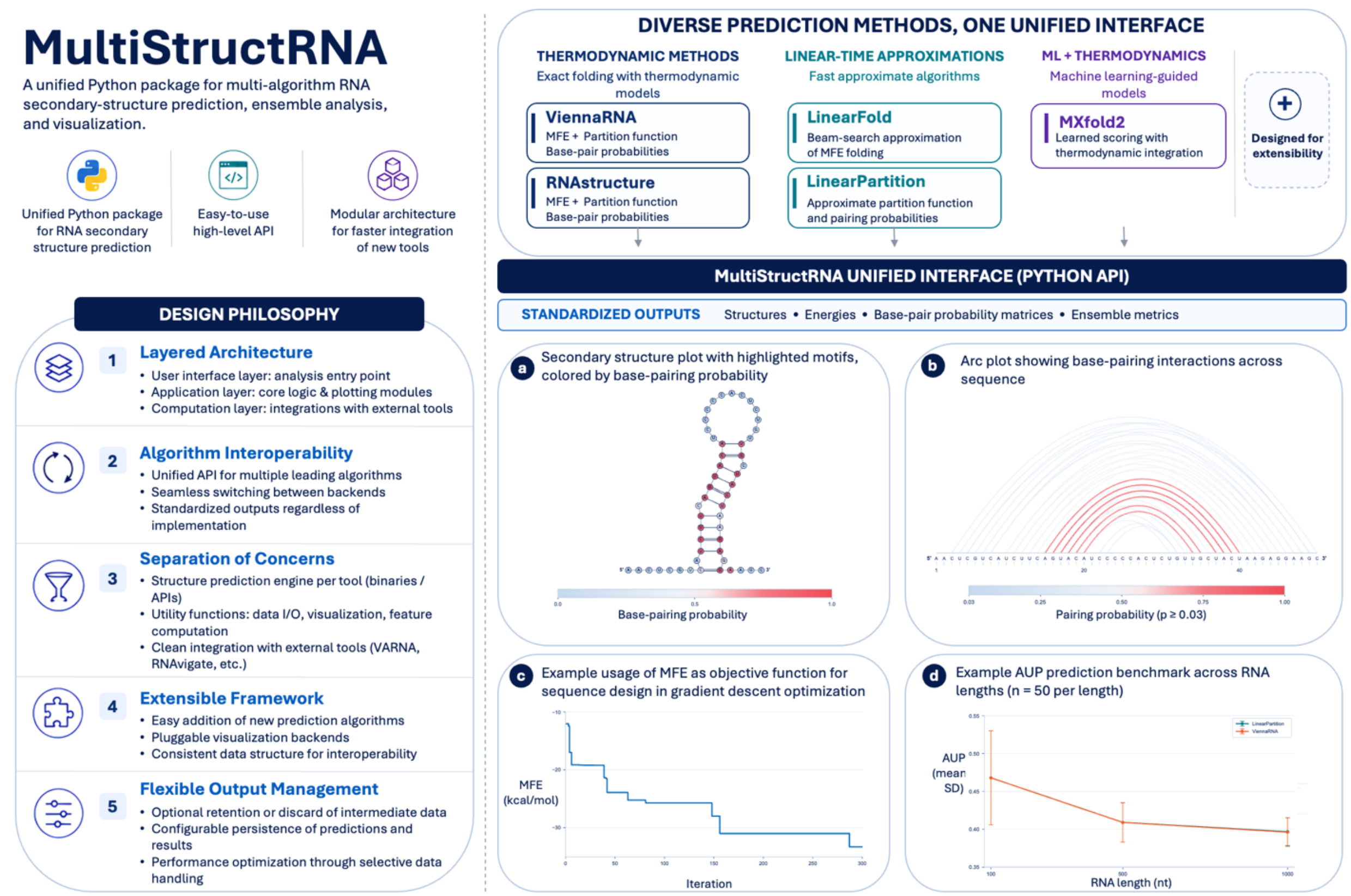

**Figure 1.** MultiStructRNA overview and example outputs. Left: design philosophy and layered architecture. Right: supported prediction backends and representative visualizations produced by MultiStructRNA, including (a) a secondary structure plot colored by base-pairing probability, (b) an arc plot showing base-pairing interactions across a sequence, (c) an example objective function (MFE – Minimum Folding Free Energy) trace during sequence optimization, and (d) an example AUP (Average Unpaired Probability) prediction benchmark across RNA lengths (n = 150 sequences per length).